\documentclass[3p,times]{elsarticle}

\usepackage{ecrc}

\volume{00}

\firstpage{1}

\journalname{Comptes Rendus Physique}

\runauth{}

\jid{procs}

\jnltitlelogo{Comptes Rendus Physique}

 \CopyrightLine{2015}{Published by Elsevier Ltd.}

\usepackage{amssymb}

\usepackage[figuresright]{rotating}

\def\XXint#1#2#3{{\setbox0=\hbox{$#1{#2#3}{\int}$}
\vcenter{\hbox{$#2#3$}}\kern-.5\wd0}}

\begin{document}

\begin{frontmatter}



\dochead{}

\title{Coarsening versus pattern formation}


\author{A. A. Nepomnyashchy}

\address{Department of Mathematics,
Technion - Israel Institute of Technology, Haifa 32000, Israel}

\begin{abstract}
It is known that similar physical systems can reveal two quite different
ways of behavior, either coarsening, which creates a uniform state or a
large-scale structure, or formation of ordered or disordered patterns,
which are never homogenized. 
We present a 
description of coarsening using simple basic models, the Allen-Cahn
equation and the Cahn-Hilliard equation, and discuss the factors that may
slow down and arrest the process of coarsening. Among them are pinning of
domain walls on inhomogeneities, oscillatory tails of domain walls,
nonlocal interactions, and others. Coarsening of pattern domains is also discussed.

\end{abstract}


\end{frontmatter}


\section{Introduction}
\label{sec1}

For many decades, the nonlinear development of instabilities in physical
systems was an object of extensive investigations. The most spectacular
consequences of instabilities are the appearance of ordered spatially
non-uniform structures ({\em pattern formation\/}, see \cite{CrHo} -
\cite{Pis}) or irregular motions ({\em spatio-temporal chaos\/}
\cite{MiLo}) under uniform external conditions. However, there is one more
scenario of an instability development: that is 
a gradual growth of the characteristic 
scale
with time
({\em coarsening\/)} 
\cite{RaVo}, \cite{Bra}.
Different evolution scenarios can take place in rather similar physical
systems.

As an example, let us consider the phase separation in binary alloys that
consist of two kinds of atoms, A and B, with volume fractions 
$\phi_A({\bf x},t)$ and
$\phi_B({\bf x},t)$, respectively. 
There exists a temperature $T_c$ such that for $T>T_c$ the components are
mixed, i.e., {\em the order parameter\/} $\phi({\bf 
x},t)=\phi_A({\bf x},t)-\phi_B({\bf x},t)$ vanishes
everywhere, while for $T<T_c$ they are separated, i.e., there exist two
thermodynamically stable phases, one with $\phi>0$ (``A-rich phase") and 
the
other with $\phi<0$ (``B-rich phase"). A mathematical model of that
phenomenon has been suggested by Cahn and Hilliard \cite{CaHi}. Under the
simplest assumption of a constant mobility, the kinetics of the phase
separation is described by the following non-dimensional equation ({\em the
Cahn-Hilliard equation\/}),
\begin{equation}
\phi_t=\nabla^2(-\phi+\phi^3-\nabla^2\phi).
\label{eq1}
\end{equation}
The uniform phase $\phi=0$ is unstable
while the uniform phases $\phi=\pm 1$ are 
stable.
The instability of the phase $\phi=0$ creates a mosaic of islands of both
stable phases.
The size of these islands ({\em 
domains\/})
grows due to
coarsening, which eventually leads to a complete separation of stable 
phases
\cite{RaVo}, \cite{Bra}.

A diblock copolymer, which consists of monomers $A$ and $B$ with reduced
equal local densities $\phi_A$ and $\phi_B$, is quite similar to a binary 
alloy.
The basic difference is the existence of a long-range interaction of 
monomers
\cite{Lei} - \cite{OoSh} which provides an additional term in the evolution
equation for the order parameter:
\begin{equation}
\phi_t=\nabla^2(-\phi+\phi^3-\nabla^2\phi)-\Gamma\phi,\;\Gamma>0,\;
\langle\phi\rangle=0.
\label{eq5}
\end{equation}
There are 
no other spatially uniform stationary 
solutions
except $\phi=0$. Therefore, when the latter solution is unstable (at 
$\Gamma<1/4$), a 
transition to a non-uniform state is unavoidable \cite{ChBr}. At small 
$\Gamma$, the 
initial evolution of disturbances is similar to that in the Cahn-Hilliard 
equation, but it is stopped when stripes with a definite pattern 
wavelength are created \cite{PoTo}.

The existence of long-wave linear instability and multiple homogeneous 
states does not guarantee the
creation of spatially uniform domains through coarsening. As  
an example, let us discuss the nonlinear dynamics governed by the 
one-dimensional {\em
Kuramoto-Sivashinsky equation\/},
\begin{equation}
\phi_t=-\phi_{xx}-\phi_{xxxx}-(\phi^2)_x,\;
\langle\phi\rangle=0,
\label{eq8}   
\end{equation}
which is used for the description of instabilities in reaction-diffusion
systems \cite{KuTs}, \cite{YaKu}, instabilities of flame fronts \cite{Siv},
and film flow instabilities \cite{Nep}. In that case, the 
linear part of the equation is identical to that of the
Cahn-Hilliard equation, and any constant solution,
$\phi=\phi_0$, satisfies equation (\ref{eq8}). However, all solutions
corresponding to uniform states are unstable. The Kuramoto-Sivashinsky
equation is a paradigmatic model of the {\em spatio-temporal chaos\/}
\cite{YaKu}; stable periodic patterns are also possible \cite{Nep},
\cite{Nep95a}, but the attraction domain of that regime is small. 
Generally, the
way of the instability development depends significantly on the details of  
the system nonlinearity and symmetry \cite{Nep95b}, \cite{CMCMV}.

Note that pattern formation and coarsening are not incompatible phenomena.   
Let us return to model (\ref{eq5}) that describes formation of stripes. 
Because of
the rotational invariance of the problem, the orientation of stripes is 
arbitrary.
Initially, a disordered system of stripes is developed from random initial  
conditions, and then 
the mean size of ordered domains grows with time, i.e., 
domain
coarsening takes place for differently oriented stripe patterns rather than 
for
different uniform phases \cite{ChBr}, \cite{BaVi}.

One can see that the interplay between coarsening and pattern formation is 
nontrivial, and it is the subject of the present chapter. Let us 
emphasize that here we discuss only {\em dynamic models\/} of 
coarsening and pattern formation, which do not include any kind of 
noise. The phenomena caused by thermal fluctuations are considered in 
other papers of the present issue and in the comprehensive book 
\cite{DeKa}, where the reader can find many additional references on 
that subject.

\section{Coarsening in one dimension: dynamics of domain walls}

When 
considering the
kinetics of coarsening, one has to take into account the following basic 
factors.

1. {\em The existence of a Lyapunov functional.\/} 
If the temporal evolution of an 
$n$-component order
parameter $\phi_i$, $i=1,\ldots,n$ is described by a {\em 
gradient\/}
evolution equation
\begin{equation}
\phi_{i,t}=-\sum_{k=1}^nD_{ik}\frac{\delta F}{\delta\phi_k},\;i=1,\ldots,n,
\label{eq8a}
\end{equation}
where $F=\int L(\phi,\phi_x,\ldots)dx$ is the 
Lyapunov functional of
the system, and $D_{ij}$ is a positive definite matrix, then
\begin{equation}
F_t=\sum_i\int\dot{\phi}_i\frac{\delta F}{\delta\phi_i}dx=-\sum_{i,k}\int    
D_{ik}\frac{\delta F}{\delta\phi_i}\frac{\delta F}{\delta\phi_k}dx\leq 0.
\label{eq8b}
\end{equation}
In the case of an equilibrium phase transition, the existence of the 
Lyapunov functional (free energy) is the consequence of the 
thermodynamics.
The nonlinear development of instabilities in systems far from 
equilibrium
in some cases is also governed, at least approximately, by {\em potential\/}
systems of equations that possess
Lyapunov functionals.

2. {\em The existence of a conservation law.\/} In the case where the order 
parameter
is a density of a conserved quantity (e.g., the number of molecules), the 
evolution
equations have a {\em divergence} form,
\begin{equation}
\phi_{i,t}=-\nabla\cdot{\bf J}_i,\;i=1,\ldots,N,
\label{eq8c}
\end{equation}
where the flux ${\bf J}_i$ is a function of the order parameter and its 
derivatives.

In order to understand how both factors influence the coarsening kinetics, 
let us   
consider a number of examples.

\subsection{Non-conserved order parameter}

\subsubsection{Allen-Cahn equation}

We start with {\em the Allen-Cahn equation\/} \cite{AlCa}
\begin{equation}
\phi_t=\phi_{xx}+\phi-\phi^3,
\label{eq9}
\end{equation}  
which describes a phase transition in the absence of a conservation law 
for the
order parameter $\phi$. 
The physical
interpretation of that model can be as follows
\cite{Bra}: $\phi$ is the spontaneous magnetization directed along the 
definite
axis
(due to
the crystal anisotropy). 
Also, the Allen-Cahn equation 
is the simplest example of an
order
parameter equation for a pattern forming system far from the thermodynamic 
equilibrium.
Let us consider the onset of
convection in a horizontal cylinder heated from
below. Above the instability threshold, the rotation of the liquid in the 
transverse section of the cylinder can be either counterclockwise 
($\phi>0$) or clockwise ($\phi<0$). The temporal evolution of the order 
parameter $\phi(x,t)$ ($x$ is the coordinate along the axis of the 
cylinder) is governed by the Allen-Cahn equation 
\cite{Nep89}. 

Equation (\ref{eq9}) has a Lyapunov functional,
\begin{equation}
F(t)=\int 
L(x,t)dx,\;L(x,t)=\frac{1}{2}\phi_x^2+\frac{1}{4}(\phi^{2}-1)^{2} \geq 
0.
\label{density}
\end{equation}
Its derivative
\begin{equation}
\dot{F}=\int\left(\frac{\partial L}{\partial\phi}\phi_t+\frac{\partial 
L}{\partial\phi_{x}}\phi_{xt}\right)dx=\int(-\phi+\phi^3-\phi_{xx})\phi_tdx=
-\int dx\phi_t^2\leq 0
\label{monotonicity}
\end{equation}
is non-positive, hence the Lyapunov functional decreases monotonically with 
time until a stationary state is reached.
It is obvious that equation (\ref{eq9}) has three fixed points
corresponding to uniform phases, $\phi=0$ (unstable paramagnetic phase) 
and $\phi=\pm 1$ (ferromagnetic phases with opposite orientations of the 
magnetization); the latter 
solutions
correspond to the absolute minima of the Lyapunov functional, $F=0$. The
separatrices
\begin{equation}
\phi_{\pm}=\pm \tanh\frac{x-\xi}{\sqrt{2}}, \;\;\xi=const
\label{sep}
\end{equation}
({\em kink} and {\em antikink}) describe {\em domain walls} separating
semi-infinite domains with different signs
of the order parameter. Note that
\begin{equation}
\phi_{\pm}\sim\pm\{1-2\exp[\mp(x-\xi)\sqrt{2}]\} \mbox{ as } 
x\to\pm\infty.
\label{eq9b}
\end{equation}
The contribution of a domain wall to the Lyapunov functional
is
$F_{0}=2\sqrt{2}/3>0$.

Other stationary solutions of (\ref{eq9})
describe {\em spatially periodic 
patterns\/}
which can be considered as periodic arrays of domains with alternating 
signs of
$\phi$. They are expressed through the elliptic Jacobi function.

Let us impose an infinitesimal 
disturbance 
$\hat{\phi}(x,t)$ on the stationary solution $\phi(x)$ and consider its 
temporal evolution governed by the 
linearized problem,
\begin{equation}
\hat{\phi}_t=\hat{\phi}_{xx}+(1-3\phi^2(x))\hat{\phi},\;|\hat{\phi}|<\infty 
\mbox{ as } x\to\pm\infty.
\label{linear}
\end{equation}
For normal disturbances $\tilde{\phi}(x)\exp(\sigma t)$ we obtain an eigenvalue 
problem,
\begin{equation}
\sigma\tilde{\phi}=\tilde{\phi}^{\prime\prime}+(1-3\phi^2)\tilde{\phi},\;\;\;
\;|\tilde{\phi}(\pm \infty|<\infty.
\label{Schroedinger}
\end{equation}

It is obvious that solution $\phi=0$ is unstable: $\sigma(\tilde{k})=1-
\tilde{k}^2$ for $\tilde{\phi}=e^{ i\tilde{k}x}$, and the solutions 
$\phi=\pm 1$
are
stable: $\sigma(\tilde{k})=-2-\tilde{k}^2$ for the same $\tilde \phi$. The 
domain
wall solutions (\ref{sep})  are neutrally stable. One can show that all the 
spatially-periodic 
solutions mentioned above are
unstable.

Let us discuss now the temporal evolution of the system, when the initial 
state is the unstable phase $\phi=0$ with a certain initial, spatial 
disordered, perturbation. According to the stability analysis presented 
above, the final state should be uniform or contain a single defect, 
a domain wall. However, it is clear that the decomposition of the phase 
$\phi=0$
can produce numerous domains
with alternating signs of $\phi$. Such a state can be characterized by a 
certain
{\em density of defects\/} decreasing with time, or by a {\em mean domain 
length\/}
which
grows with time. 
At late stages of the evolution,
when the typical distance between domain walls is large 
the analysis can be done by means of asymptotic methods
\cite{Gorshkovetal74},
\cite{Kawasaki_Ohta82}.
One considers a set of 
domain
walls (\ref{sep}) of alternating signs centered at $\xi_{i}=\xi_{i}(t)$, 
$i=1,2,
\ldots,$ and slowly moving because of their interaction.
By means of asymptotic expansions, the original nonlinear problem is 
transformed into an infinite system of inhomogeneous linear equations. Their 
solvability conditions determine the 
following equations of motion for the centers of 
domain walls \cite{Nep10}:
\begin{equation}
\frac{2\sqrt{2}}{3} \dot{\xi}_{i}=-\frac{\partial U}{\partial 
\xi_{i}},
\label{eqmot}
\end{equation}
where
\begin{equation}
U=\sum_{i}W(\xi_{i}-\xi_{i-1}),\;W(\xi_{i}-\xi_{i-1})=
-8\sqrt{2}\exp\left[-\sqrt{2}(\xi_{i}-\xi_{i-1})\right].
\label{partintenergy}
\end{equation}
Thus, the domain walls attract to each other according to the exponential 
interaction law (\ref{eqmot}) which reflects the exponential domain 
wall
asymptotics (\ref{eq9b}).

System (\ref{eqmot}) has a family of stationary solutions
\begin{equation}
\xi_{j}=a+jl,\;\;\; j=0,\pm
1,\pm 2,\ldots \label{per}
\end{equation}
corresponding to periodic patterns 
with the spatial period $2l$. 
The attractive interaction makes all these solutions
unstable.
%

Consider the
interaction
of a pair of domain walls.
Setting $l(t)=\xi_{2}(t)-\xi_{1}(t)$, we find that the distance between 
domain
walls is
governed by the equation
\begin{equation}
\frac{2\sqrt{2}}{3}l^{\prime}(t)=-32e^{-l\sqrt{2}}.
\label{eqdist}
\end{equation}
If $l(0)=l_{0}\gg 1$, the solution is
\begin{equation}
l=l_{0}+\frac{1}{\sqrt{2}}\ln\left(1-48e^{-\sqrt{2}l_{0}}t\right).
\label{dist1}
\end{equation}
The distance between two domain walls becomes of $O(1)$ at
\begin{equation}
t\sim t_{0}=\frac{1}{48}e^{\sqrt{2}l_{0}}+O(1).
\label{time}
\end{equation}
Finally, the domain walls reach the distance of order $O(1)$ and 
annihilate.
During the time interval $t$,
only the domain
walls with the original separation greater than $\sim \ln(48t)/ \sqrt{2}$ 
can
survive. Thus, a logarithmic coarsening takes place. In a large but finite 
system with the length $L$, the number of domain walls $N(t)\leq L/l(t)$.

As an example of the situation when two locally stable phases are 
energetically
non-equivalent, let us consider a system with the Lyapunov functional
\begin{equation}
F[\phi(x)]=\int\left[\frac{1}{2}\phi_x^2+\frac{1}{4}(1-\phi^2)^2-h\phi\right]dx
\label{eq2.17}
\end{equation}
which corresponds to the dynamic equation
\begin{equation}
\phi_t=\phi_{xx}+\phi-\phi^3+h.
\label{eq2.17a}
\end{equation}
In the case of a magnetic system, the last term in the expression 
(\ref{eq2.17}) describes the influence of 
an
external magnetic field, which makes the orientation of the magnetization 
in 
the
direction of the field preferable. The uniform stationary states satisfy 
the
equation
\begin{equation}
\phi-\phi^3+h=0.
\label{eq2.18}
\end{equation}  
In the interval $-h_*<h<h_*$, 
equation (\ref{eq2.18}) has three
solutions:
stable solution $\phi=\phi_+>1/\sqrt{3}$, another stable solution
$\phi=\phi_-<-1/\sqrt{3}$,
and
an intermediate unstable solution $\phi_0$, 
$-1/\sqrt{3}<\phi_0<1/\sqrt{3}$. For
$h>0$,
the phase with $\phi=\phi_+>0$ is stable, and the phase with $\phi=\phi_-$ 
is
metastable.

If $h$ is small, domain walls are described by formulas (\ref{sep}) at the 
leading
order. By means of an asymptotic analysis, one can find that
the motion of each domain wall is governed by the equation \cite{Nep10}
\begin{equation}
\frac{2\sqrt{2}}{3}\frac{d\xi}{dt}=\mp 2h
\label{eq9a}
\end{equation}  
(the upper sign is for a kink and the lower sign is for an antikink).
This motion creates a coarsening process, which is significantly faster than 
that in the case of energetically equivalent phases. That process leads to 
the elimination of the
metastable phase.

\subsubsection{Fractional Allen-Cahn equation}

The logarithmic law of the one-dimensional coarsening is caused by the 
exponentially weak interaction, due to the exponential asymptotics (\ref{eq9b}) 
characteristic for solutions of partial differential equations. However, the 
basic equations governing the natural phenomena are often integro-differential 
equations rather than partial differential equations. For instance, the free 
energy density of a fluid depends on the fluid density in a nonlocal way 
\cite{Pis02}. The conventional van der Waals' expression for the fluid free 
energy which contains a squared density gradient \cite{vdW} corresponds to a 
certain asymptotic limit of the basic nonlocal expression. The 
asymptotics of a domain wall (i.e., the gas-liquid boundary) in the framework 
of local and nonlocal models are strongly different: while a local model 
predicts an exponential decay, a nonlocal model suggests a power-law decay 
\cite{Pis02}. Fronts that 
cannot be governed by local partial differential equations have been found also 
in studies of transitions with long range
interactions \cite{Hei}, vacancy diffusion and domain growth in binary
alloys
\cite{FVP}, and ordering kinetic on fractal structures \cite{MBM} (for the 
latter subject, see \cite{BCV}).
As an example let 
us discuss the front propagation in systems with {\em superdiffusion\/} 
\cite{MeKl}. While the normal diffusion is associated with a Gaussian 
non-correlated random walk of particles, the superdiffusion is observed in 
non-equilibrium systems with an algebraically decaying jump length 
distribution, where the central limit theorem is no valid.
Among the examples are wave turbulence \cite{Bal}, transport in porous 
media
\cite{GKS}, and forage trajectories of animals \cite{VABMPS}. 
One 
can use a
superdiffusive generalization of the Allen-Cahn equation, 
\begin{equation}
\phi_t=D_{|x|}^{\gamma}\phi+\phi-\phi^3,\;1<\gamma<2.
\label{eq9c}
\end{equation}
Here $D_{|x|}^{\gamma}$ denotes the fractional Riesz derivative, which can be 
defined by its action in the Fourier space:
\begin{equation}
F\left(D_{|x|}^{\gamma}\phi(x)\right)(k)=-k^{\gamma}F\left(\phi(x)\right)(k),
\label{eq9d}  
\end{equation}
where $F$ is the symbol of the Fourier transform.
Equation (\ref{eq9c}) has a Lyapunov 
functional
\cite{NNG}. In a contradistinction to the exponential asymptotics 
(\ref{eq9b}) of
the domain walls in the {\em local\/} PDE (\ref{eq9}), the domain walls 
in the
integro-differential equation (\ref{eq9c}) has an algebraic tail,
\begin{equation}
\phi_+\sim 1-\frac{\sec(\pi\gamma/2)}{2\Gamma(2-\gamma)}x^{-\gamma}, \; 
x\gg 1.
\label{eq9e}
\end{equation}
That leads to a power law for the kink-antikink attraction,
\begin{equation}
l^{\prime}(t)=-Cl^{-\gamma}
\label{eq9f}
\end{equation}
(cf. (\ref{eqdist})), and for the temporal decay of the number of domain 
walls on a
finite spatial interval, $N\sim t^{-1/\gamma}$ \cite{NNG}.

\subsubsection{Non-potential systems}

For systems far from equilibrium, e.g., in the case of longwave 
instabilities of flows,
the Lyapunov functional generally does not exist. Nevertheless, coarsening 
may take
place if the domain walls are described by monotonic functions. As an example, let us 
mention the amplitude equation that governs 
the fixed-flux
convection in a tilted slot \cite{CeYo}:
\begin{equation}
\phi_t=\phi_{xx}+\phi-\phi^3+2\alpha\phi\phi_x, \; \alpha>0.
\label{eq10}
\end{equation}
Because the symmetry $x\to -x$ is violated, the kink and antikink 
domain walls
have different widths:
\begin{equation}
\phi_{\pm}(x)=\tanh\beta_{\pm}(x-\xi),\;\beta_{\pm}=\frac{1}{2}(\alpha\pm\sqrt{\alpha^2+2}).
\label{eq11}
\end{equation}
The interaction of domain walls is attractive but asymmetric. For a pair 
which consists
of a kink with the center in the point $\xi_1(t)$ and an antikink with the 
center in
the point $\xi_2(t)$, the equations of motion look as
\begin{equation}
\xi_1^{\prime}=F_+\exp(-2|\beta_-|(\xi_2-\xi_1)),\;
\xi_2^{\prime}=F_-\exp(-2\beta_+(\xi_2-\xi_1)),
\label{eq12}
\end{equation}  
where $F_+\neq F_-$. All the periodic stationary solutions (patterns) are 
unstable, and
a logarithmically slow coarsening takes place, similarly to the case of the 
Allen-Cahn
dynamics.

\subsection{Conserved order parameter}

\subsubsection{Potential systems}

Let us return to the Cahn-Hilliard equation (\ref{eq1}), which can be 
written as
\begin{equation}
\phi_t+j_x=0,\;j=(\phi_{xx}+\phi-\phi^3)_x;\;-l\leq x\leq l.
\label{eq13}
\end{equation}
Being a generic nonlinear equation governing longwave instabilities in the 
presence of the conservation law \cite{Nep95b}, 
that equation has been revealed in 
numerous problems of different physical nature, including secondary flows produced by 
the instability of the Kolmogorov flow \cite{Nep76}, Marangoni instability of a 
two-layer system with a deformable interface \cite{SiNe}, nonlinear development of 
zigzag instability of convection rolls \cite{MNT}, and even coarsening of ordered 
domains in oscillatory patterns governed by the complex Swift-Hohenberg equation 
\cite{GeKn}, which describes oscillations in lasers \cite{JMN94}, \cite{JMN95} and 
optical parametric oscillations \cite{LoGe} - \cite{SSW}. 
  
For sake of simplicity, apply the boundary conditions $\phi_x=\phi_{xxx}=0$ at 
$x=\pm l$; then
\begin{equation}
j(\pm l)=0,\; \frac{d}{dt}\int_{-l}^l\phi dx=0
\label{eq14}
\end{equation}
(the total length of domains of 
each phase is conserved).
The Lyapunov functional (\ref{density}) decreases with time:
\begin{equation}
F_t=-\int (\phi_{xx}+\phi-\phi^3)^2 dx\leq 0.
\label{eq15}  
\end{equation}
Using the boundary conditions, one can present equation (\ref{eq13}) in
the form \cite{Kawasaki_Ohta82}
\begin{equation}
\partial_x^{-2}\phi_t+\phi_{xx}+\phi-\phi^3+h(t)=0,\;
\partial_x^{-2}\phi_t(x,t)=\frac{1}{2}\int_{-l}^l|x-y|\phi_t(y,t)dy.
\label{eq16}
\end{equation}
Note that despite the energetical
equivalence of phases
$\phi=\pm 1$, an efficient field $h(t)$ appears, which must to be determined 
self-consistently.
The motion of a kink and 
an antikink towards
each other would change the total lengths of domains of different phases, 
and hence it is not
possible. Two kinks can move simultaneously towards an antikink placed 
between them, or
kink-antikink pairs can move as a whole. The correlated motion of $n$ 
domain walls with the centers at
$\xi_i$, $i=1,\ldots,n$, sufficiently far from each other, is governed by the system 
\cite{Kawasaki_Ohta82} 
\begin{equation}
-\sum_{j=1}^n2(-1)^{i-j}|\xi_i-\xi_j|\xi_j^{\prime}=16\sum_{j\neq
i}e^{-|\xi_i-\xi_j|\sqrt{2}}\mbox{sign}(\xi_j-\xi_i)+2(-1)^ih(t),\;i=1,\ldots,n,
\label{eq18}
\end{equation}
supplemented by the conservation law
\begin{equation}
\sum_{i=1}^n(-1)^i\xi_i^{\prime}=0.
\label{eq19}
\end{equation}
For a kink-antikink pair, the attraction is compensated by the field
$h=8\exp[-(\xi_2-\xi_i)\sqrt{2}]$, hence the domain walls are motionless. 
For a symmetric
kink-antikink-kink triplet with coordinates of domain walls $\xi_1=-l(t)$, 
$\xi_2=0$ and
$\xi_3=l(t)$, one obtains $h=0$, $l^{\prime}=-8\exp(-l\sqrt{2})$, hence the 
annihilation time
depends on $l_0=l(0)$ as
$$t_0\sim\frac{1}{8\sqrt{2}}l_0e^{l_0\sqrt{2}}$$
(cf. (\ref{time})).

\subsubsection{Non-potential systems}

As an example of a non-potential system with a conservation law let us 
consider {\em the
convective Cahn-Hilliard equation\/},
\begin{equation}
\phi_t+(\phi_{xx}+\phi-\phi^3)_{xx}-\frac{D}{2}(\phi^2)_x=0,\;-\infty<x<\infty,
\label{eq20}
\end{equation}
which has been suggested to describe several physical processes,
namely spinodal decomposition of phase separating systems in an external 
field
\cite{Leu} - \cite{EmBr}, step instability on a crystal surface
\cite{SaUw}, faceting of growing, thermodynamically unstable surfaces
\cite{LiMe} - \cite{WORD}, evolving nanofoams \cite{CCGV} as well as 
dewetting of a thin film flowing
down an inclined plane \cite{Thiele}. That equation provides ``a bridge" 
between the
Cahn-Hilliard equation (\ref{eq1}) ($D=0$) and the Kuramoto-Sivashinsky 
equation
(\ref{eq8}) ($\phi\to
-2\phi/D$, $D\gg 1$). 

Stationary patterns
$\phi=\phi(x)$ are described by the problem
\begin{equation}
\phi^{\prime\prime\prime}+(\phi-\phi^3)^{\prime}-\frac{D\phi^2}{2}=-\frac{DA}{2},\;
-\infty<x<\infty;\;A>0;\;x\to\pm\infty,\;|\phi|<\infty.
\label{eq21}
\end{equation}
For any $D\neq 0$, the set of solutions of equation (\ref{eq21}) is 
incomparably more complex than that of the usual Cahn-Hilliard equation. One 
can easily find some exact solutions of the problem. The constant solutions,
\begin{equation}   
\phi=\phi_{\pm}=\pm\sqrt{A},
\label{eq22}
\end{equation}
correspond to two stable phases. For domain walls,
there exist {\em exact\/} solutions \cite{Leu},
one for a kink with
$A=A_+=1+D/\sqrt{2}$,
\begin{equation}
\phi=\phi_+(x)=\phi_+^0\tanh\frac{\phi_+^0}{\sqrt{2}}(x-\xi),\;
\phi_+^0=\sqrt{1+D/\sqrt{2}},\;\; \xi=const,
\label{eq25}
\end{equation}
and the other for an antikink with
$A=A_-=1-D/\sqrt{2}$, $D<\sqrt{2}$,
\begin{equation}
\phi=\phi_-(x)=-\phi_-^0\tanh\frac{\phi_-^0}{\sqrt{2}}(x-\xi),\;
\phi_-^0=\sqrt{1-D/\sqrt{2}},\;\; \xi=const.
\label{eq26}
\end{equation}
However, the set of stationary solutions is much more rich. Specifically,
solution (\ref{eq25}) is just one representative of a {\em family\/} of kinks
$\phi_+(x;A)$. In addition to the monotonic antikink (\ref{eq26}), there 
exists also a discrete set of non-monotonic antikink solutions \cite{ZPNG} 
(that phenomenon is typical for models containing higher-order spatial 
derivatives, see \cite{PeTr}- \cite{LPP}).  The 
kink-antikink pair is formed by antikink (\ref{eq26}) and a 
representative of the family of kinks with $A=A_-$.

If  
$0<D<D_0=\sqrt{2}/3$. In that region,  the coarsening is observed \cite{EmBr}, 
\cite{GDN98}, \cite{GDN99}, \cite{WORD}.
Because of the asymmetry between kinks and antikinks, a kink-antikink pair 
moves  
spontaneously with a definite velocity $v_2(D,L)$. The most typical process 
observed by 
coarsening is the annihilation of domain wall triplets, when two kinks of 
the same sign
approach with velocities $\pm v_3(D,L)$ the kink of the opposite sign 
situated between
them. Exact expressions for $v_2(D,L)$ and $v_3(D,L)$ can be found in 
\cite{PZRGN}. In
the limit of small $D$ \cite{WORD},
$$v_2(D,L)\sim v_3(D,L)\sim -(D^2\sqrt{2}/4)\exp(-DL/2).$$
Therefore, the coarsening law is logarithmic.

For $D>D_0$, the domain walls have oscillatory tails. That case will be 
discussed in the
next section.   

\section{Factors hindering coarsening}

In the present section, we discuss some typical situations where the system 
cannot reach a
uniform state or another energetically preferred state by coarsening.

\subsection{External inhomogeneities}

The motion of domain walls leading to annihilation can be stopped by 
inhomogeneity of the
medium. Recall that we consider the phenomena in the absence of noise. 
Coarsening in inhomogeneous systems in the presence of thermal 
fluctuations is considered in \cite{Cor}.
 
In a potential system, the domain wall would ``prefer" the location 
where its
energy will be smaller than in other locations. For example, let us 
consider 
the following 
modification
of the one-dimensional Allen-Cahn equation \cite{Nep89}:
\begin{equation}
\phi_t=\phi_{xx}+[1+\epsilon f(x)]\phi-\phi^3.
\label{ampinhom}
\end{equation}
At the leading order in $\epsilon$, the equation of motion for a domain 
wall of any kind
is
\begin{equation}
\frac{2\sqrt{2}}{3}\frac{d\xi}{dt}=-\frac{d}{d\xi}V_{ih}(\xi),\mbox{ where }
V_{ih}(\xi)=\frac{1}{2}\int_{-\infty}^{\infty}f(\xi+y)\cosh^{-2}(y)dy.\;
\label{eq2.22}
\end{equation}
Specifically, if the inhomogeneity has a $\delta$-like shape,
$$f(x)=-2V_0\delta(x-x_*),$$
the interaction potential is
$$V_{ih}(x_0)=-V_0\cosh^{-2}\frac{x_0-x_*}{\sqrt{2}}.$$
Generally, the shape of the potential is a linear transformation of the inhomogeneity shape, 
according to (\ref{eq2.22}).

If there are many domain walls and many inhomogeneities, the motion of 
domain walls
is determined by the system of equations (\ref{eqmot}) with the potential
$$U=\sum_{i}W(\xi_i-\xi_{i-1})+\sum_{i}V_{ih}(x_i),$$
where $W(\xi_i-\xi_{i-1})$ is determined by equation (\ref{partintenergy}), 
and
$V_{ih}(x_i)$ corresponds to (\ref{eq2.22}). Thus, the problem of
finding stationary solutions of (\ref{ampinhom}) is equivalent to finding 
equilibrium
configurations of a chain of particles in the external potential
(\ref{eq2.22}),
interacting according to the law (\ref{partintenergy}).
This model resembles the
well-known Frenkel--Kontorova model (see e.g. \cite{Peyrard_Aubry83}). 

As an example let us consider two domain walls with coordinates $\xi_1$ and 
$\xi_2$
which are near the distant $\delta$-shaped attracting inhomogeneities:
$$f(x)=-2V_0\delta(x-x_{1*})-2V_0\delta(x-x_{2*}),\;V_0>0,\;x_{2*}>x_{1*}.$$
$x_{2*}-x_{1*}=l_*\gg 1$, $|\xi_1-x_{1*}|=O(1)$, $|\xi_2-x_{2*}|=O(1)$.
The equation of motion for the left domain wall is:
\begin{equation}
\frac{2\sqrt{2}}{3}\frac{d\xi_1}{dt}=16e^{-(\xi_2-\xi_1)\sqrt{2}}
-\sqrt{2}V_0\sinh\frac{\xi_1-\xi_{1*}}{\sqrt{2}}\cosh^{-3}\frac{\xi_1-x_{1*}}{\sqrt{2}}.
\label{eq2.23}
\end{equation}
The first term in the
right-hand side of
equation can be estimated as $16\exp(-l_*\sqrt{2})$. The minimum of the
second term in
the right-hand side of
the equation is equal to $-2\sqrt{2}V_0/3\sqrt{3}$. Thus, we come to the
conclusion that if
\begin{equation}
\frac{2\sqrt{2}}{3\sqrt{3}}V_0>16e^{-l_*\sqrt{2}},
\label{eq2.24}
\end{equation}
the domain wall will not be able to escape from the potential well created 
by
the
inhomogeneity. Hence, the coarsening will be stopped when the distances 
between
the neighbor domain walls satisfy the inequality (\ref{eq2.24}).

Similarly, in the case of domain walls pushed by the asymmetry of phases 
(see (\ref{eq9a}), we
find the criterion of pinning:
$$V_0>\frac{3\sqrt{3}}{\sqrt{2}}h.$$

If $f(x)$ is a periodic function, the sequence of pinning sites (minima of the 
potential) filled by pinned domain walls can be regular (``commensurate patterns") 
\cite{Cou} or
irregular (``spatial chaos") \cite{CER}.

\subsection{Oscillatory tails of domain walls and stability of stationary patterns}

In the examples considered in Sec. 2, the domain walls are described by 
monotonic functions like (\ref{sep}). The monotonicity
of the asymptotic behavior of the domain wall solution on the
infinity leads to a sign-preserving (attracting) interaction
between domain walls. Oscillatory tails of domain walls create
a sign-alternating  interaction potential.
The domain walls can be
captured near the potential minima, therefore stable patterns are formed 
due to pinning of a domain wall by an
inhomogeneity created by another domain wall.
   
As an example, let us consider the stability of periodic solutions of the convective
Cahn-Hilliard equation (\ref{eq20}), which satisfy the condition $\phi(x+l)=\phi(x)$. At large $l$, 
these solutions resemble periodic sequences of domain walls. Define the 
pattern wavenumber $K=2\pi/l$. The normal 
disturbances of a periodic solution have the shape of a Floquet-Bloch function, 
$\hat{\phi}(x,t)=\tilde{\phi}(x)\exp(ikx+\sigma t)$, where $\tilde{\phi}(x+l)=\tilde{\phi}(x)$, 
and $k$ is a quasi-wavenumber, $|k|<K/2$. A periodic solution is always neutrally stable 
($\sigma=0$) with respect to a spatial shift, $\tilde{\phi}(x)=\phi_x(x)$, $k=0$. Therefore, a 
special attention has to be payed to potentially unstable longwave disturbances with small $k$. 
Their growth rate $\sigma(k;K)$ can be presented as
$$\sigma(k;K)=\sigma_1(K)k+\sigma_2(K)k^2+\ldots$$
One can show that the sign of $\sigma_1^2(K)$ depends on the dependence of the squared 
pattern amplitude
$$A=\frac{1}{l}\int_0^l\phi^2(x)dx$$
(see (\ref{eq21})) on the pattern wavenumber $K=2\pi/l$ 
\cite{Nep}, \cite{PoMi}, \cite{ZPNG}. If $dA/dK<0$ for any $K$, which takes place for 
$D<D_0=\sqrt{2}/3$, then $\sigma_1^2(K)>0$ for any $K$, therefore all periodic solutions are 
unstable. That is compatible with the attractive interaction between domain walls. For 
$D>D_0$, the function $A(K)$ is not monotonic, and the extrema of $A(K)$ separate the regions 
of a monotonic growth of longwave disturbances, $\sigma_1^2(K)>0$, and those of the 
oscillatory response of patterns to dilations and compressions, $\sigma_1^2(K)<0$ \cite{FST}.
The region
of oscillatory response can contain a subinterval of stable patterns (where $\sigma_2(K)<0$) 
\cite{GNDZ},
\cite{ZPNG}. That 
is possible because of the alternating sign of the interaction between domain walls.
On the boundaries of the stability interval, the patterns 
become unstable with respect to either longwave (phase) disturbances (see 
\cite{MiPo}, \cite{NMP}) or shortwave disturbances with $k=K/2$, leading to a 
spatial period doubling.

Because the potential of the interaction between domain walls has multiple
minima, the distance between domain walls is not selected in a unique way.
The pattern includes elements
with ``short" and ``long" distances between the maxima that alternate in a
rather irregular way \cite{GNDZ}.

\subsection{Nonlocal interaction}

A specific kind of an arrested coarsening process has been found for 
equation (\ref{eq5}) \cite{PoTo}, 
which can be written also as
\begin{equation}
\partial_x^{-2}(\phi_t+\Gamma\phi)+\phi_{xx}+\phi-\phi^3+h(t)=0.
\label{eq26a}   
\end{equation}
For small $\Gamma$
the lowest density of the 
Lyapunov functional is achieved for patterns with 
wavelength $\lambda_{opt}=O(\Gamma^{-1/3})$. According to the linear 
stability theory, the disturbance with largest growth rate has a 
wavelength $\lambda_c=O(1)$. Therefore, one could expect that the 
energetically preferable, long-wave pattern will be developed from the initial short-wave 
pattern 
by coarsening. The coarsening process takes place indeed, but it is stopped when the wavelength 
reaches a much smaller value, $\lambda_{min}=O(\ln(1/\Gamma))$. The criterion of the pattern 
stabilization is similar to (\ref{eq2.24}), but now the stabilizing factor is the 
nonlocal 
interaction which is proportional to $\Gamma$. 
 
\subsection{Pattern-induced pinning of a domain wall.}

The stability regions of a periodic pattern and a uniform state may
overlap. In that case, the behavior of a {\em domain wall between the
pattern and the uniform state\/} is crucial. 
Near the threshold of the instability creating short-wave patterns, where the width of a 
domain wall is large compared to the
characteristic pattern wavelength, one can describe the dynamics of a
domain wall using the envelope function approach \cite{NeWh}, \cite{Seg}.
In the framework of that approach one comes to the conclusion that a
domain wall between the
pattern and the uniform state moves with a constant velocity, which is proportional to the 
difference between the Lyapunov functional densities of
the phases \cite{MNT}. However, the 
influence of the underlying periodic pattern leads to some qualitative changes of the domain 
wall dynamics. 
First, 
the 
motion of the domain wall is an oscillatory process; during one 
period, one stripe is created or melted \cite{AMPT}, \cite{Sch}. 
Secondly,    
because of the pinning effect, there is a finite interval of the   
parameter value where the domain wall is motionless, i.e., a pattern and
a uniform state coexist. Near the threshold of the pattern appearance,
that interval is transcendentally small \cite{Pom}, \cite{MNT}.

As an example, let us consider the competition and coexistence between 
patterns and uniform
states for a system governed by the Swift-Hohenberg equation,
\begin{equation}
\phi_t=\left[\epsilon-\left(\frac{\partial^2}{\partial
x^2}+1\right)^2\right]\phi-\phi^3
\label{eq27}
\end{equation}
which corresponds to the Lyapunov functional
\begin{equation}
F\{\phi\}=\int\left\{-\frac{\epsilon}{2}\phi^2+\frac{1}{4}\phi^4
+\frac{1}{2}\left[\left(\frac{\partial^2}{\partial
x^2}+1\right)\phi\right]^2\right\}dx.
\label{eq28}
\end{equation}
That model was suggested for studying hydrodynamics fluctuations near
the instability threshold \cite{SwHo} and used for modelling B\'{e}nard
convection \cite{GrCr}.
At $\epsilon>0$, periodic patterns exist
with wavenumbers $k$ in the interval
$1-\sqrt{\epsilon}<k^2<1+\sqrt{\epsilon}$, and they are stable in a
subinterval $k_-(\epsilon)<k<k_+(\epsilon)$. At
$\epsilon>1$, constant nonzero solutions
$\phi_{\pm}=\pm\sqrt{\epsilon-1}$ appear. At $\epsilon>3/2$ they become
stable with respect to small disturbances; kinks with oscillatory tails
connect both
stable uniform phases $\phi_{\pm}$ 
\cite{OuFu}.

The value of the Lyapunov functional density for the regular pattern with
an optimal wavenumber is lower than that of the uniform state when
$\epsilon<\epsilon_m\approx 6.3$ \cite{OuFu}, \cite{AMPT}. Nevertheless,
the domain wall between both states is immobile for much smaller values of
$\epsilon$, $\epsilon>\epsilon_c\approx 1.7574$. The reason is the
self-induced pinning caused by the oscillatory asymptotic perturbation of
the uniform state. Similarly, the pinning effect prevents the replacement
of a pattern by a uniform state at $\epsilon>\epsilon_m$. The stability
interval for a finite fragment of patterns sandwiched between
semi-infinite regions of a uniform state slightly depends on the length of
that fragment \cite{AMPT}. Note that the noise activates the transition
for a metastable state to a truly stable, energetically preferred, state
\cite{AMPT}.

The coexistence of patterns and uniform states has been revealed for many
pattern-forming systems (for a review, see \cite{Kno}).

\section{Coarsening in two and three dimensions: curvature effects}

\subsection{Phase separation}

First, let us consider domain coarsening for spatially uniform
states in a system without the conservation law.

In a two-dimensional (three-dimensional) potential system, the Lyapunov functional
can be diminished without annihilation of a domain wall, just by
diminishing its length (area). Let us consider a two-dimensional Allen-Cahn
equation,
\begin{equation}
\phi_t=\phi_{xx}+\phi_{yy}+\phi-\phi^3+h,
\label{eq29}
\end{equation}
and present the isoline $\phi(x,y,t)=0$ (``a front"), which describes the
center of a curved domain walls between the stable uniform phases, in the 
form $y=H(x,t)$: $\phi(x,y,t)<0$ as $y<H(x,t)$, $\phi(x,y,t)>0$ as
$y>H(x,t)$. One can show that the motion of the domain wall is determined,
in the limit of small $h$ and a small curvature of the front, by the
equation \cite{PeLi}, \cite{Nep10},
\begin{equation}
\frac{H_t}{\sqrt{1+H_x^2}}=\frac{H_{xx}}{(1+H_x^2)^{3/2}}
-\frac{3\sqrt{2}}{2}h, \mbox{ or }
v=\kappa-\frac{3\sqrt{2}}{2}h,
\label{eq3.3}
\end{equation}
where $v$ is the normal velocity, and $\kappa$ is the curvature of
the front.
Specifically, in the case $h=0$ (both phases have the same free energy), we
get just the relation $v=\kappa$,
which is called {\em curvature flow\/}.

As an example, let us consider a round droplet of the phase $\phi_-$ in the
infinite sea of the phase $\phi_+$. Because $\kappa=1/R$, in the
case $h=0$ the
droplet radius is changed according to the law
$$R^2(t)=R(0)^2-2t.$$
The droplet collapses during the finite time $t_*\approx R_0^2/2$.
The
obtained life time of the droplet shows that the characteristic coarsening
scaling is $l\sim O(t^{1/2})$, which is significantly faster than in the
one-dimensional case. The same coarsening law is obtained in the 3D case.

Moreover, even for the fractional Allen-Cahn equation,
\begin{equation}
\phi_t=-(-\nabla^2)^{\gamma/2}\phi+\phi-\phi^3+h,\;1<\gamma<2,
\label{eq31}
\end{equation}
the front motion is determined by a formula similar to (\ref{eq3.3}) (up to
numerical coefficients that depend on $\gamma$), and the scaling $l\sim
O(t^{1/2})$ is established on the late stage of coarsening (at the initial
state, $l\sim
O(t^{1/\gamma})$, according to the scaling properties of the linearized
equation) \cite{NNG}.

A more significant change of the front dynamics is produced by {\em memory\/}, 
when the temporal evolution of the order parameter is governed by the equation
\begin{equation}
\phi_t=-\int_0^t a(t-s)\frac{\delta F}{\delta\phi}(x,s)ds.
\label{eq31a}
\end{equation}
Equation (\ref{eq3.3}) is replaced by 
\begin{equation}
\frac{v_t}{1-\alpha v^2}+\gamma v=\kappa-\frac{3\sqrt{2}}{2}h(1-\alpha 
v^2)^{1/2},
\label{eq31b}
\end{equation}
where the constants $\alpha$ and $\gamma$ are determined by the Laplace transform 
of the kernel $a(t-s)$ \cite{RoNe}, \cite{RDN}.

In the presence of a conservation law, using the Cahn-Hilliard equation, one can 
find that
the evaporation of a single round droplet of the phase $\phi_-$ in the infinite 
sea of the  
phase $\phi_+$ is governed by the equation \cite{Bra}
$$R^3(t)=R^3(0)-\frac{3}{2}\sigma t,$$
where $\sigma$ is a parameter corresponding to the effective surface tension of 
the domain
wall. Hence, the scaling law $l\sim O(t^{1/3})$ is predicted. In the case of 
droplets of
different sizes,
 the main mechanism of coarsening is the
growth of big droplets (with a
smaller curvature) at the expense of small droplets (with a larger
curvature), which
leads to ``flattening" of the interphase boundary and hence the decrease of
the
Lyapunov functional (``Ostwald ripening").
Lifshitz
and
Slyozov \cite{LiSl} and Wagner \cite{Wag} have developed a kinetic theory of the 
phase
separation in the limit of small concentration of the minority phase. A 
detailed description of that theory and its extensions can be found in 
\cite{DeKa}. Here we mention some basic results. 
There exists a critical radius $R_c(t)\sim t^{1/3}$ such that smaller 
droplets 
evaporate by
diffusion, while larger droplets grow by absorbing the matter through the 
majority phase.
They have obtained a self-similar droplet radius distribution and found the law 
$R(t)\sim
t^{1/3}$ for the characteristic droplet radius. Later, the latter law was 
confirmed,
theoretically and numerically, for arbitrary concentrations of phases
\cite{Hus}-\cite{TCG}. Note that at shorter time after the beginning of the 
phase separation
process, a scaling law $R(t)\sim t^{1/4}$ has been predicted and observed
\cite{Maz}-\cite{Sza}. The same scaling laws are observed in the framework of a 
more general
model,
\begin{equation}
\phi_t=-\nabla\cdot[M(\phi)(\nabla^2\phi+\phi-\phi^3)]
\label{eq31c}
\end{equation}
with a non-constant mobility function $M(\phi)$. For instance, the crossover 
from $R(t)\sim
t^{1/4}$ to $R(t)\sim
t^{1/3}$ has been observed for $M(\phi)=1-\phi^2$ \cite{PBL}.

A nontrivial kind of Cahn-Hilliard equation has been derived for the
description of flows in thin liquid films in the presence of disjoining pressure
\cite{ODB}-\cite{NSL}. Here the conservation law is the the conservation of the liquid
volume, while ``the phases" are a macroscopic film and a mesoscopic ``precursor film".
The Cahn-Hilliard equation describes the decomposition of a film into droplets connected by 
the
thin precursor film. The coarsening of droplets is due to the growth of large droplets
at the expense of small ones and because of the motion of droplets
leading to their coalescence. In the framework of the standard model \cite{GlWi}, the
coarsening law is $N(t)\sim t^{-2/5}$, where $N(t)$ is the number of
droplets. Generalizations of that model leading to different coarsening rates can be
found in \cite{Kit}.

A specific kind of coarsening takes place if there are more than two
thermodynamically equivalent phases \cite{Lif}, \cite{Saf}, e.g., because 
of
different possible orientations of the spin (the list of examples can be 
found in
\cite{Saf}). The problem can be modelled by means of an {\em overdamped 
sine-Gordon
equation\/} similar the Allen-Cahn equation (\ref{eq9}) but with the 
potential
$V(\phi)=\cos p\phi-1$; the stable equilibrium phases correspond to the 
potential
maxima, $\phi=2\pi m/p$, $m=0,1,\ldots,p-1$.
For $p=2$, the coarsening is similar to that for the Allen-Cahn equation in 
any
dimension. If $p>2$, the coarsening in 1D is determined by the 
exponentially weak
interaction of domain walls, which can be now either attractive or 
repulsive. For
$p\geq d+1$, a logarithmic rate of coarsening is also predicted \cite{Saf}.

Coarsening in non-potential systems was studied using isotropic \cite{CCGV} and 
anisotropic \cite{GDN99} generalizations of the convective Cahn-Hilliard equation.

A numerical analysis of coarsening versus pattern formation in non-potential 
pattern-forming systems has been carried out in \cite{GoPi}.

\subsection{Pattern ordering}

The naturally appearing patterns usually contain numerous defects 
\cite{NPL}.
Specifically,
the   
pattern may have a multidomain structure.
One can distinguish between two kinds of domain walls in periodic patterns
\cite{MNT}. The first kind of domain walls separates patterns of different
symmetry, which are generally not energetically equivalent (e.g., stripe 
patterns
and hexagonal patterns). In that case, the domain wall tends to move 
expanding the
energetically prefered domain, but it can be stopped by pinning, as it was
explained in the previous sections \cite{MNT}. The second kind of domain 
walls 
separates patterns of the
same type but with different orientations or different values of the 
wavelength. 
Domains of different orientations appear spontaneously or are created by the side 
walls in a finite region \cite{TrKu}.
Different scenarios of domain wall evolution are possible \cite{MNT}: (i) the domain 
wall can
be a source of
a wavenumber selection, similarly to a side wall
\cite{CDHS} or a ramp smoothly matching pattern region with a subcritical 
region
\cite{KBBC}; (ii) the domain wall can be destroyed by an intrinsic 
instability;
(iii) it can spread and smooth down.

Besides domain walls, patterns contain dislocations \cite{MeNe}, 
\cite{NePi},
coupled pairs of dislocations \cite{PiNe}, \cite{CNL}, and 
disclinations \cite{PaNe} 
-  
\cite{GoNe}. Their motion is also a significant factor of the ordering in 
periodic
patterns \cite{MNT}, \cite{CNL}.

While some specific phenomena related to the dynamics of defects in 
patterns have
been a subject of a theoretical analysis, the full picture of pattern 
ordering is
studied mostly by means of numerical simulations of dynamical equations 
(possibly with noise) or experimentally.

In isotropic systems with symmetry $\phi\to -\phi$ (e.g., for (\ref{eq5}) 
or (\ref{eq27})), 
stripes of different orientations are generated. That allows a simplified 
description of patterns by the phase field $\phi({\bf x},t)$, $\phi({\bf 
x},t)=\phi_0\cos[\psi({\bf x},t)]$, with the local wavevector ${\bf 
k}({\bf x},t)=\nabla\psi({\bf x},t)$ \cite{PoMa}. The numerical 
simulations reveal power laws for the growth of domains and elimination of 
domain walls, 
dislocations and disclinations for the original equation (\ref{eq27}) 
\cite{QiMa03} and for the phase equation \cite{QiMa04}. Computations 
carried out for (\ref{eq5}) show that on the background of the final 
state, which is a unidomain 
structure, the orientational two-time correlation function has a 
power-law asymptotics, while the spatial two-point correlation function 
is 
subject to a transition from a power law to an exponential law with time 
\cite{RRM}. 

If the symmetry $\phi\to -\phi$ is violated (due to an external field 
\cite{Jag} or a cubic term in the free energy density \cite{OhSh}), a 
competition between stripes and hexagons takes place. As an example, let 
us mention 
ordering in patterns governed by a generalized 
Swift-Hohenberg
equation
\begin{equation}
\phi_t=-\frac{\delta F}{\delta\phi},\;
F\{\phi\}=\int\left\{-\frac{\epsilon}{2}\phi^2+\frac{1}{4}\phi^4+\frac{s}{3}\phi^3
+\frac{1}{2}[(\nabla^2+1)\phi]^2\right\}d{\bf x},\;{\bf x}=(x,y).
\label{eq32}
\end{equation}
Changing $s$, one can arrange a transition between stripes and hexagons and 
vice versa.
The analysis of the transition has been done \cite{OhSh} by studying the 
structure
factor $S({\bf k}, t)=\langle|\hat{\phi}({\bf k},t)|^2\rangle$, where 
$\hat{\phi}({\bf
k},t)$ is the Fourier transform of the order parameter $\phi({\bf x},t)$, 
and  
$\langle\rangle$ denotes ensemble averaging. One has found different 
scaling
laws for ordering the stripes, for 
the 
growth of
hexagonal domains due to the stripe-to-hexagon transition, and 
for the
growth of stripes from a disordered hexagonal patterns.

Orientational ordering in hexagonal patterns has been studied experimentally in 
\cite{HATC} and numerically in \cite{GVV}, using the modification of equation (\ref{eq5}) 
with 
broken inversion theory. Note that the problem of the orientational ordering is related 
to the problem of coarsening in the system with degenerate phases studied in \cite{Lif}, 
\cite{Saf}.

In conclusion, we have reviewed basic models and effects related to coarsening in 
pattern forming systems. Recently, investigations of more complex cases have been 
initiated, e.g., domain coarsening in an oscillatory patterns  
\cite{MML}, pattern coarsening in time-dependent domains \cite{KnKr}, and patterns in 
networks \cite{ImSh}. These subjects are beyond the scope of the present 
review.





\bibliographystyle{elsarticle-num}
\bibliography{<your-bib-database>}

\begin{thebibliography}{1}

\bibitem{CrHo}
M. C. Cross and P. C. Hohenberg, Rev. Mod. Phys. {\bf 65}, 851 (1993).

\bibitem{Hoy}
R. Hoyle, {\it Pattern Formation: An Introduction to Methods}, Cambridge
University Press, 2006.

\bibitem{Pis}
L. M. Pismen, {\it Patterns and Interfaces in Dissipative Dynamics},
Springer, Berlin, 2006.

\bibitem{MiLo}
A. S. Mikhailov and A. Yu. Loskutov, {\it Foundations of Synergetics II:
Chaos and Noise}, Springer, Berlin, 2012.

\bibitem{RaVo}
L. Ratke and P. W. Voorhees, {\it Growth and Coarsening: Ostwald Ripening 
in
Materials Processing}, Springer, Berlin, 2002.

\bibitem{Bra}
A. J. Bray, Adv. Phys. {\bf 43}, 357 (1994).

\bibitem{CaHi}
J. W. Cahn and J. E. Hilliard, J. Chem. Phys. {\bf 28}, 258 (1958).

\bibitem{Lei}
L. Leibler, Macromolecules {\bf 13}, 1602 (1980).

\bibitem{OhKa}
T. Ohta and K. Kawasaki, Macromolecules {\bf 19}, 2621 (1986).

\bibitem{OoSh}
Y. Oono and Y. Shiwa, Mod. Phys. Lett. B {\bf 1}, 49 (1987).

\bibitem{ChBr}
J. J. Christensen and A. J. Bray, Phys. Rev. E {\bf 58}, 5364 (1998).

\bibitem{PoTo}
P.Politi snd A. Torcini, arXiv: 1412.6269v1 [cond-mat.stat-mech] (2014).

\bibitem{KuTs}
Y. Kuramoto and T. Tsuzuki, Prog. Theor. Phys. {\bf 55}, 356 (1976).

\bibitem{YaKu}
T. Yamada and Y. Kuramoto, Prog. Theor. Phys. {\bf 56}, 681 (1976).

\bibitem{Siv}
G. I. Sivashinsky, Acta Astronautica {\bf 4}, 1177 (1977).

\bibitem{Nep}
A. A. Nepomnyashchy, Fluid Dynamics {\bf 9}, 354 (1974).

\bibitem{Nep95a}
A. A. Nepomnyashchy, Europhys. Lett. {\bf 31}, 437 (1995).

\bibitem{Nep95b}
A. A. Nepomnyashchy, Physica D {\bf 86}, 90 (1995).

\bibitem{CMCMV}
M. Castro, J. Mu\~{n}os-Garc\'{i}a, R. Cuerno, M. M. 
Garc\'{i}a-Hern\'{a}ndez, and L. V\'{a}zques, New J. of Phys. {\bf 9}, 102 
(2007).

\bibitem{BaVi}
D. Bayer and J. Vi\~{n}als, Phys. Rev. E {\bf 65}, 046119 (2002).

\bibitem{DeKa} 
R. C. Desai and R. Kapral, {\it Dynamics of Self-Organized and
Self-Assembled
Structures}, Cambridge University Press, 2009.

\bibitem{AlCa}
S. M. Allen and J. W. Cahn, Acta Met. {\bf 20}, 423 (1972).

\bibitem{Nep89}
A. A. Nepomnyashchy, in {\it Nonlinear waves 1,
     Dynamics and Evolution}, ed. by A. V. Gaponov-Grekhov,  M. I.
Rabinovich,
     and J. Engelbrecht, 103 (Springer-Verlag, Berlin et al., 1989).

\bibitem{Gorshkovetal74}
K. A. Gorshkov, L. A. Ostrovsky, and E. N. Pelinovsky, Proc. IEEE {\bf 62}, 
1511
(1974).

\bibitem{Kawasaki_Ohta82}
K. Kawasaki and T. Ohta, Physica A {\bf 116}, 573 (1982).

\bibitem{Nep10}
A. A. Nepomnyashchy, in {\it Pattern Formation at Interfaces, CISM Courses
and Lectures, vol. 513}, ed. by P. Colinet and A. Nepomnyashchy, 59
(Springer-Verlag, Wien, New York, 2010).

\bibitem{Pis02}
L. M. Pismen, Colloids and Surfaces A {\bf 206}, 11 (2002).

\bibitem{vdW}
J. D. van der Waals, Z. Phys. Chem. {\bf 13}, 657 (1894). English translation: 
J. S. Rowlinson, J. Stat. Phys. {\bf 20}, 197 (1979).

\bibitem{Hei}
H. Heinrich, J. Stat. Mech. Theory Exp., P07006 (2007).

\bibitem{FVP}
C. Frontera, E. Vives, and A. Planes, Phys. Rev. B {\bf 48}, 9321 (1993).

\bibitem{MBM}
U. Marini Bettolo Marconi, Phys. Rev. E {\bf 57}, 1290 (1998).

\bibitem{BCV}
R. Burioni, F. Corberi, and A. Vezzani, J. Stat. Mech. Theory Exp., 
P02040
(2009); P12024 (2010).

\bibitem{MeKl}
R. Metzler and J. Klafter, Phys. Rep. {\bf 339}, 1 (2000).

\bibitem{Bal}
A. M. Balk, J. Fluid Mech. {\bf 467}, 163 (2002).

\bibitem{GKS}
D. N. Gerasimov, V. A. Kondratieva, and O. A. Sinkevich, Physica D {\bf 
239}, 1593
(2010).

\bibitem{VABMPS}
G. M. Viswanathan, V. Afanasyev, S. V. Buldyrev, E. J. Murphy, P. A. 
Prince, and H.
E. Stanley, Nature {\bf 381}, 413 (1996).

\bibitem{NNG}
Y. Nec, A. A. Nepomnyashchy, and A. A. Golovin,  Physica D {\bf 237}, 3237 
(2008).

\bibitem{CeYo}
P. Cessi and W. R. Young, J. Fluid Mech. {\bf 237}, 57 (1992).
   
\bibitem{Nep76}
A. A. Nepomnyashchy, J. Appl. Math. Mech. {\bf 40}, 836 (1976). 

\bibitem{SiNe}
I. B. Simanovskii and A. A. Nepomnyashchy, {\it Convective Instabilities in Systems with
Interface}, Gordon and Breach, Singapore, 1993.

\bibitem{MNT}
B. A. Malomed, A. A. Nepomnyashchy, and M. I. Tribelsky, Phys.
Rev. A {\bf 42}, 7244 (1990).

\bibitem{GeKn}
L. Gelens and E. Knobloch, Phys. Rev. B {\bf 80}, 046221 (2009).

\bibitem{JMN94}
J. Lega, J. V. Moloney, and A. C. Newell, Phys. Rev. Lett. {\bf 73}, 2978 (1994).

\bibitem{JMN95}
J. Lega, J. V. Moloney, and A. C. Newell, Physica D {\bf 83}, 478 (1995).

\bibitem{LoGe}
S. Longhi and A. Geraci, Phys.
Rev. A {\bf 54}, 4581 (1996).

\bibitem{SMRVS}
V. J. S\'{a}nchez-Morcillo, E. Rold\'{a}n, G. J. de Valc\'{a}rel, and K. Staliunas, 
Phys.
Rev. A {\bf 56}, 3237 (1997).

\bibitem{SSW}
K. Staliunas, G. Slekys, and C. O. Weiss, Phys. Rev. Lett. {\bf 79}, 2658 (1997).
  
\bibitem{ODB}
A. Oron, S. H. Davis, and S. G. Bankoff, Rev. Mod. Phys. {\bf 69}, 931 (1997).

\bibitem{GlWi}
K. B. Glassner and T. P. Witelski, Phys. Rev. E {\bf 67}, 016302 (2003).

\bibitem{SVR}
V. M. Starov, M. G. Velarde, and C. J. Radke, {\it Wetting and Spreading Dynamics}, CRC 
Press, Boca Raton, 2007.

\bibitem{NSL}
A. Nepomnyashchy, I. Simanovskii, and J. C. Legros, {\it Interfacial Convection in 
Multilayer Systems}, Second Edition, Springer, New York {\em et al.\/}, 2012.

\bibitem{Kit}
G. Kitavtsev, Eur. J. Appl. Math. {\bf 25}, 83 (2014).

\bibitem{Leu}
K. Leung,
J. Stat. Phys. {\bf 61}, 345 (1990).

\bibitem{YRHJ}
C. Yeung, T. Rogers, A. Hernandes-Machado, and D. Jasnow,
J. Stat. Phys. {\bf 66}, 1071 (1992).

\bibitem{EmBr}
C. L. Emmott and A. J. Bray,
Phys. Rev. E {\bf 54}, 4568 (1996).

\bibitem{SaUw}
Y. Saito and M. Uwaha,
J. Phys. Soc. Jpn. {\bf 65}, 3576 (1996).

\bibitem{LiMe}
F. Liu and H. Metiu,
Phys. Rev. B {\bf 48}, 5808 (1993).

\bibitem{GDN98}
A. A. Golovin, S. H. Davis, and A. A. Nepomnyashchy,
Physica D {\bf 122}, 202 (1998).

\bibitem{GDN99}
A. A. Golovin, S. H. Davis, and A. A. Nepomnyashchy,
Phys. Rev. E {\bf 59}, 803 (1999).

\bibitem{GNDZ}
A. A. Golovin, A. A. Nepomnyashchy, S. H. Davis, and M. A. Zaks,
Phys. Rev. Lett. {\bf 86}, 1550 (2001).

\bibitem{WORD}
S. J. Watson, F. Otto, B. Y. Rubinstein, and S.H. Davis,
Physica D {\bf 178}, 127 (2003).

\bibitem{CCGV}
M. Castro, R. Cuerno, M. M. Garc\'{i}a-Hern\'{a}ndez, and L. V\'{a}zques, 
Phys. Rev. Lett. {\bf 112}, 094103 (2014).

\bibitem{Thiele}
U. Thiele, M. G. Velarde, K. Neuffer, M. Bestehorn, and Y. Pomeau,
Phys. Rev. E {\bf 64}, 061601 (2001).

\bibitem{ZPNG}
M. A. Zaks, A. Podolny, A. A. Nepomnyashchy, and A. A. Golovin, SIAM J.
Appl. Math. {\bf 66}, 700 (2006).

\bibitem{PeTr}
L. A. Peletier and W. C. Troy, {\it Spatial Patterns: Higher Order Models 
in Physics and Mechanics}, Birkh\"{a}user, Boston, 2001.

\bibitem{NeVe}
V. I. Nekorkin and M. G. Velarde, {\it Synergetic Phenomena in Active 
Lattices: Patterns, Waves, Solitons, Chaos}, Springer, Berlin, 2002.

\bibitem{LPP}
Th. Le Goff, P. Politi, and O. Pierre-Louis, Phys. Rev. E {\bf 90}, 032114 
(2014). 

\bibitem{PZRGN}
A. Podolny, M. A. Zaks, B. Y. Rubinstein, A. A. Golovin, and A. A. 
Nepomnyashchy, Physica
D {\bf 201}, 291 (2005).

\bibitem{Cor}
F. Corberi, in this issue.

\bibitem{Peyrard_Aubry83}
M. Peyrard and S. Aubry, J. Phys. C {\bf 16}, 1593 (1983).

\bibitem{Cou}
P. Coullet, Phys. Rev. Lett. {\bf 56}, 724 (1986).

\bibitem{CER}  
P. Coullet, C. Elphick, and D. Repaux, Phys. Rev. Lett. {\bf 58}, 431 
(1987).

\bibitem{PoMi}
P. Politi and Ch. Misbah, Phys. Rev. Lett. {\bf 92}, 090601 (2004).

\bibitem{FST}
U. Frisch, Z. S. She, and O. Thual, J. Fluid Mech. {\bf 168}, 221 (1986).

\bibitem{MiPo}
Ch. Misbah and P. Politi, Phys. Rev. E {\bf 80}, 030106 (R) (2009).

\bibitem{NMP}
M. Nicoli, Ch. Misbah, and P. Politi, Phys. Rev. E {\bf 87}, 063302 
(2013).

\bibitem{NeWh}
A. C. Newell and J. A. Whitehead, J. Fluid Mech. {\bf 38}, 279 (1969).

\bibitem{Seg}
L. A. Segel, J. Fluid Mech. {\bf 38}, 203 (1969).

\bibitem{AMPT}
I. S. Aranson, B. A. Malomed, L. M. Pismen, and L. S. Tsimring, 
Phys.
Rev. E {\bf 62}, R5 (2000).

\bibitem{Sch}
A. Scheel, Arch. Rational Mech. Anal. {\bf 181}, 505 (2006).

\bibitem{Pom}
Y. Pomeau, Physica D {\bf 23}, 3 (1986).

\bibitem{SwHo}
J. Swift and P. C. Hohenberg, Phys.
Rev. A {\bf 15}, 319 (1977)

\bibitem{GrCr}
H. S. Greenside and M. C. Cross, Phys.
Rev. A {\bf 31}, 2492 (1985).



\bibitem{OuFu}
K. Ouchi and H. Fujisaka, Phys.
Rev. E {\bf 54}, 3895 (1996).

\bibitem{Kno}
E. Knobloch, Nonlinearity {\bf 21}, T45 (2008).

\bibitem{PeLi}
{\it Dynamics of Curved Fronts}, ed. by P. Pelce and A. Libchaber, 
Academic Press, Boston {\em et al.\/}, 1988.

\bibitem{RoNe}
H. G. Rotstein and A. A. Nepomnyashchy, Physica D {\bf 136}, 245 (2000).

\bibitem{RDN}
H. G. Rotstein, A. I. Domoshnitsky, and A. A. Nepomnyashchy, Physica D {\bf 146}, 
137 (2000).

\bibitem{LiSl}
I. M. Lifshitz and V. V. Slyozov, J. Phys. Chem. Solids {\bf 19}, 35 
(1961).

\bibitem{Wag}
C. Wagner, Z. Elektrochem. {\bf 65}, 581 (1961).
   
\bibitem{Hus}
D. A. Huse, Phys. Rev. B {\bf 34}, 7845 (1986).

\bibitem{ASM}
J. G. Amar, F. E. Sullivan, and R. D. Mountain, Phys. Rev. B {\bf 37}, 196 
(1988).
   
\bibitem{RED}
T. M. Rogers, K. R. Elder, and R. C. Desai,  Phys. Rev. B {\bf 37}, 9638 (1988).

\bibitem{TCG}
R. Toral, A. Chakrabarti, and J. D. Gunton, Physica A {\bf 213}, 41 (1995).

\bibitem{Maz}
G. F. Mazenko, Phys. Rev. B {\bf 49}, 5747 (1991).

\bibitem{CCZ} 
F. Corberi, A. Coniglio, and M. Zannetti, Phys. Rev. E {\bf 51}, 5469 (1995).

\bibitem{Sza}
G. Szab\'{o}, Phys. Rev. E {\bf 57}, 6172 (1998).

\bibitem{PBL}
S. Puri, A.J. Bray, and J. L. Lebovitz, Phys. Rev. E {\bf 56}, 758 (1997).

\bibitem{Lif}
I. M. Lifshitz, Sov. Phys. JETP {\bf 15}, 939 (1962).

\bibitem{Saf}
S. A. Safran, Phys. Rev. Lett. {\bf 46}, 1581 (1981).

\bibitem{GoPi}
A. A. Golovin and L. M. Pismen, Chaos {\bf 14}, 845 (2004).

\bibitem{NPL}
A. C. Newell, T. Passot, and J. Lega, Annu. Rev. Fluid Mech. {\bf 25}, 184 
(1993).

\bibitem{TrKu}
A. K. Tripathi and D. Kumar, Phys. Rev. E {\bf 90}, 022915 (2014).

\bibitem{CDHS}
M. C. Cross, P. G. Daniels, P. C. Hohenberg, and E. D. Siggia, J. Fluid 
Mech. {\bf
127}, 155 (1983).

\bibitem{KBBC}
L. Kramer, E. Ben-Jacob, H. Brand, and M. C. Cross, Phys. Rev. Lett. {\bf 
49}, 1891
(1981).

\bibitem{MeNe}
D. Meiron and A. C. Newell, Phys. Lett. A {\bf 113}, 289 (1985).

\bibitem{NePi}
A. A. Nepomnyashchy and L. M. Pismen, Phys. Lett. A {\bf 153}, 427 (1991).

\bibitem{PiNe}
L. M. Pismen and A. A. Nepomnyashchy, Europhys. Lett. {\bf 24}, 461 (1993).

\bibitem{CNL}
P. Colinet, A. A. Nepomnyashchy, and J. C. Legros, Europhys. Lett. {\bf 
57}, 480
(2002).

\bibitem{PaNe}
T. Passot and A. C. Newell, Physica D {\bf 74}, 301 (1994).

\bibitem{BoNe}
C. Bowman and A. C. Newell, Rev. Mod. Phys. {\bf 70}, 289 (1998).

\bibitem{GoNe}
A. A. Golovin and A. A. Nepomnyashchy, Phys. Rev. E {\bf 67}, 056202 
(2003).

\bibitem{PoMa}
Y. Pomeau and P. Manneville, J. Phys. (France) Lett. {\bf 40}, L-609 
(1979).

\bibitem{QiMa03}
H. Qian and G. F. Mazenko, Phys. Rev. E {\bf 67}, 036102 (2003).

\bibitem{QiMa04}
H. Qian and G. F. Mazenko, Phys. Rev. E {\bf 69}, 011104 (2004).

\bibitem{RRM}
Ch. Riesch, G. Radons, and R. Magerle, Phys. Rev. E {\bf 90}, 052101 
(2014).

\bibitem{Jag}
E. A. Jagla, Phys. Rev. E {\bf 70}, 046204 (2004).

\bibitem{OhSh}
H. Ohnogi and Y. Shiwa, Phys. Rev. E {\bf 84}, 011611 (2011).

\bibitem{HATC}
C. Harrison, D. E. Angelescu, M. Trawick, Zh. Chang {\em et al.\/}, Europhys. Lett. 
{\bf 67}, 800 (2004).

\bibitem{GVV}
L. R. G\'{o}mez, E. M. Vall\'{e}s, and D. A. Vega, Phys. Rev. Lett. {\bf 97}, 
188302 (2006). 

\bibitem{MML}
B. Marts, K. Martinez, and A. L. Lin, Phys. Rev. E {\bf 70}, 056223 (2004).

\bibitem{KnKr}
E. Knobloch and R. Krechetnikov, J. Nonlin. Sci. {\bf 24}, 493 (2014).

\bibitem{ImSh}
R. Imayama and Y. Shiwa, Phys. Rev. E {\bf 80}, 036117 (2009).

\end{thebibliography}







\end{document}